\documentclass{iopart}
\usepackage{graphicx}
\usepackage{amssymb}

\begin{document}

\title[Vortex Shear Banding Transitions in Superconductors with Inhomogeneous...]{Vortex Shear Banding Transitions in Superconductors with Inhomogeneous Pinning Arrays}
\author{C. Reichhardt and  C. J. O. Reichhardt}
\address{Theoretical Division and Center for Nonlinear Studies,
  Los Alamos National Laboratory, Los Alamos, New Mexico 87545, USA
}
\ead{cjrx@lanl.gov}

\begin{abstract}
  We numerically examine the flow of superconducting vortices in samples containing square pinning arrays in which a band of pins is removed.  When a drive is applied at an angle with respect to the band orientation, we find that the vortex depinning initiates in the pin-free channel.  The moving vortices form a series of quasi-one-dimensional shear bands that begin flowing in the bulk of the pin-free channel, and the motion gradually approaches the edge of the pinned region.  The consecutive depinning of each shear band appears as a series of jumps in the velocity-force curves and as sharp steps in the spatial velocity profiles.  When a constant drive is applied parallel to the pin-free channel along with a gradually increasing perpendicular drive, the net vortex velocity decreases in a series of steps that correspond to the immobilization of bands of vortices, and in some cases the flow can drop to zero, creating a field effect transistor phenomenon.  These results should also be relevant to other types of systems that exhibit depinning in the presence of inhomogeneous pinning.

\end{abstract}
\maketitle

\vskip 2pc

\section{Introduction}
When a solid is subjected to a shear force, the dynamic response
can be characterized by constructing a stress-strain curve.
As the strain increases from zero,
the system remains in a solid state with a linearly growing stress, and 
when the strain is large enough,
the system yields and starts to flow 
as bonds between the constituent particles break.
In many cases, for strains above yielding, the stress saturates
or even decreases with increasing strain \cite{Beer09,Argon79,Regev13}.
Yielding
can occur via
shear banding or shear localization, in which
only a portion of the sample is moving while
other portions remain stationary,
or it can occur
via the development of a smooth velocity gradient throughout the sample
\cite{Cohen06,AlixWilliams18}. 
The behavior at yielding varies depending on the rate of change of the strain
as well as
on whether
the system is crystalline, polycrystalline, or amorphous.

Another phenomenon that has similarities to the yielding transition is
the depinning transition for
a collection of particles driven over a random or periodic substrate
\cite{Fisher98,Reichhardt17}.
Under an increasing drive,
the system can remain
pinned until a critical drive is reached above which flow initiates.
The motion can be elastic, with each particle maintaining the same neighbors,
or plastic, with continuous rearrangements of the particles.
When the substrate is composed of pinning sites, individual particles can
be trapped directly at the pinning sites or indirectly through interactions with
pinned particles.
For solids undergoing yielding, it is only particle-particle
interactions that produce resistance to flow,
and in materials with a higher shear modulus,
a larger force must be applied in order to
induce yielding. 
In some cases, systems that undergo depinning can also
exhibit a yielding transition, such as
vortices in type-II superconductors in a
Corbino geometry, which exerts a
nonuniform force on the vortex lattice
\cite{Lopez99,Benetatos02,Miguel03,Furukawa06,Okuma09,Cabral16}.
If the vortices remain elastic and rotate as a solid object,
the velocity of an individual vortex increases linearly with
its distance $r$ from the center of the disk.
If plastic flow occurs instead,
the vortices disorder, exhibit liquid behavior, and
have a velocity that
decays as  $1/r$.
Both of these behaviors have been observed in
experiment \cite{Lopez99} and simulations \cite{Benetatos02}.
The simulations indicate that there is
a critical current or rotation rate required to induce a
transition to the plastic flow state, which is accompanied by the proliferation of 
topological defects in the vortex structure \cite{Benetatos02,Miguel11}.
Since the 
vortices form a triangular lattice in the absence of pinning,
shear banding can also occur
in which a band of vortices flow together as a rigid body at one velocity
while other bands of vortices flow at different velocities,
and there can be locking and unlocking transitions
between the different moving bands due to the spatial periodicity of the vortices.
Simulations by
Lin {\it et al.} showed dynamical locking and dynamical commensurability effects for
vortices in mesoscale Corbino disks \cite{Lin09}. 
Experiments by Okuma {\it et al.} for vortices in a Corbino geometry
with an additional ac drive
produced Shapiro steps, 
interpreted as resulting from
a shear banding effect in which vortices near the center of the sample 
depin first, followed by laminar flow 
at higher drives \cite{Okuma09,Kawamura15}.
There are also studies of shearing effects for vortices interacting
with pinning sites inside a Corbino geometry \cite{Rosen13}
or with a single easy flow channel embedded 
in strong pinning \cite{Reichhardt10}.
For
inhomogeneous pinning,
composed of
a region of strong pinning coexisting 
with a channel of weak or no pinning,
the vortices in the weakly pinned channel move first
under an applied drive, forming
a velocity gradient with the fastest flow
at the center of the weakly pinned channel and
zero flow at the edge of the more strongly pinned region
\cite{Marchetti99,Marchetti00}. 
Another particle-based system in which similar shear effects can occur
is a rotating ring of trapped colloidal
particles embedded in a background of free particles,
where the
relative motion of the background particles exterior or interior to the
moving ring can be measured.
Experiments performed with this geometry also 
reveal laning or shear banding effects
\cite{Buttinoni15,Williams16,OrtizAmbriz18}.    

In this work we use numerical simulations to show
that vortex shear banding can
be achieved without a Corbino geometry
and with a uniform drive 
when the substrate consists of
a strip of strong pinning
coexisting with a pin-free channel.
We focus on 
a square pinning array in which a group of adjacent rows of pins have been
removed.
Such a substrate could be created
experimentally using artificial pinning structures
\cite{Baert95,Martin97,Trastoy14,Wang13},
patterned irradiation \cite{Banerjee03},
or controlled thickness modulations
\cite{Besseling99,Guillamon14,Dobrovolskiy19,Dobrovolskiy17,Dobrovolskiy16}.
The vortices in the pin-free channel easily depin as an elastic solid
under a parallel applied driving force.
When the drive is applied at an angle with respect to the pin-free channel,
we find
strong shear banding 
effects,
with flow initially occurring
only at the center of the pin-free channel and the vortices near the edges of
the channel remaining immobile.
As the drive increases, 
a series of shear banding transitions
occur as additional rows 
of vortices in the pin-free channel
depin, producing signature jumps in the velocity-force curves 
and steps
in the spatial velocity profile.   
Under a constant parallel drive and gradually increasing perpendicular drive,
the longitudinal velocity passes through a
series of drops as 
the shear banding effects become stronger
until there is a transition to
a state in which the vortices flow in both the parallel and perpendicular directions.
Interestingly,
the flow of vortices within the
pinned portion of the sample in this case
is strictly
in the perpendicular direction due to a dynamical symmetry locking with the square
pinning array.  
Under some conditions,
we find that application of 
a perpendicular drive can send the longitudinal velocity sharply to zero,
similar to a field effect transistor in which the current response
changes sharply under an applied field.
Our results could be tested in superconductors with artificial pinning arrays
where a portion of the pinning sites have been removed, in samples
with a step in the sample thickness,
or in systems with a gradient in the pinning density or strength
\cite{Wang13,Ray13,Guenon13}.  
Although our results are for a vortex system, they should be general
to other particle like systems
where spatially inhomogeneous pinning can arise, such as skyrmions in chiral magnets and
colloidal systems.    

\section{Simulation}

\begin{figure}
  \begin{center}
    \includegraphics[width=0.8\columnwidth]{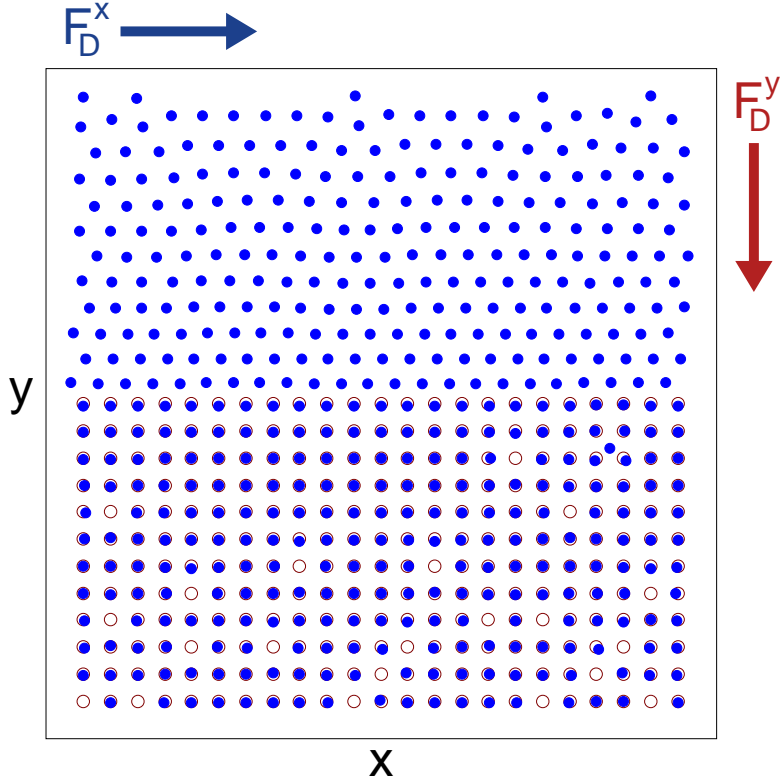}
  \end{center}
  \caption{Illustration of the sample geometry.  Pins (open circles) are placed in
    a square array, and half of the pinning sites are removed (top portion of sample).
    Vortex positions (dots) are obtained through simulated annealing and drives are
    applied parallel, $F^x_D$ (blue arrow), and perpendicular, $F^y_D$ (red arrow),
    to the $x$ axis.
  }
\label{fig:1}
\end{figure}

We consider a superconducting  system with a
square pinning lattice of lattice constant $a$ where half
the pinning sites are removed and $N_v$ vortices are present,
as illustrated in Fig.~\ref{fig:1}.
We use a rigid vortex approximation in which the dynamics 
of vortex $i$ is governed by the following overdamped equation of 
motion:
\begin{equation}  
\eta \frac{d {\bf R}_{i}}{dt} = 
{\bf F}^{vv}_{i} + {\bf F}^{vp}_{i} + {\bf F}^{x}_{D} + {\bf F}^{y}_{D} 
\end{equation}
The damping constant is $\eta$, which we set equal to unity. The
vortex-vortex interactions are repulsive and
the sum of the vortex-vortex forces on vortex $i$ is  
${\bf F}^{vv}_{i} = \sum_{j\neq i}F_{0}K_{1}(R_{ij}/\lambda){\hat {\bf R}}_{ij}$.
Here $K_{1}$ is the modified Bessel function which decays monotonically,
$R_{ij} = |{\bf R}_{i} - {\bf R}_{j}|$ is the distance between vortices $i$ and $j$,
${\hat {\bf R}_{ij}} = ({\bf R}_{i} - {\bf R}_{j})/R_{ij}$,
$\lambda$ is the penetration depth,
$F_{0} = \phi^{2}_{0}/(2\pi\mu_{0}\lambda^3)$,
$\phi_{0}$ is the flux quantum, and $\mu_{0}$ is the permittivity.
The pinning sites are modeled as parabolic traps 
of radius $R_{p} = 0.35\lambda$
and force
${\bf F}^{vp }_{i} =\sum_{k}(F_{p}R^{(p)}_{ik}/r_{p})\Theta(r_{p} -R^{(p)}_{ik}){\hat {\bf R}^{(p)}}_{ik}$.
Here $\Theta$ is the Heaviside step function,
$F_p$ is the maximum force exerted by the pinning site,
${\bf R}_k^{(p)}$ is the location of pinning site $k$,
$R_{ik}^{(p)} = |{\bf R}_{i} - {\bf R}_{k}^{(p)}|$, and
$ {\hat {\bf R}_{ik}^{(p)}} = ({\bf R}_{i} - {\bf R}_{k}^{(p)})/R_{ik}^{(p)}$.
All forces are measured in units of $F_{0}$ and lengths in units of $\lambda$.
We choose the vortex density such that the number of vortices would match
the number of pinning sites if the entire pinning lattice were present,
$B/B_{\phi}^{\rm complete}=1.0$, where $B_{\phi}^{\rm complete}$ is the matching field
for the full pinning array.
Since half of the pinning sites are missing, our vortex density
is $B/B_{\phi}=2.0$, where $B_{\phi}$ is the matching field for the half pinning array.
We set the pinning density to $B_{\phi}^{\rm complete}=0.4/\lambda^2$.
The net drive ${\bf F}_D$ is applied at an angle to the $x$ axis and
is given by ${\bf F}_D=F_D^x{\bf \hat x} - F_D^y{\bf \hat y}$.

The initial vortex positions are obtained using
a simulated annealing protocol in which we start at a high temperature
with the vortices in a liquid state and slowly cool the sample to $T=0$.
After the initialization, we apply a driving 
force that is increased in increments of $\Delta F_{D}$.
We spend 20000 simulation time steps at each increment to ensure that the
system has reached a steady state, and we measure the average vortex
velocities in the parallel,
$\langle V_x\rangle=N_v^{-1}\sum_i^{N_v}{\bf v}_i \cdot {\bf \hat x}$,
and perpendicular,
$\langle V_y\rangle=N_v^{-1}\sum_i^{N_v}{\bf v}_i \cdot {\bf \hat y}$,
directions.
We focus on the case $F_{p} = 0.75$ and
consider a range of drives over which
depinning of the vortices trapped at the pinning sites is negligible.
This model
has previously been used
to study vortex pinning and dynamics in systems with uniform
pinning or conformal pinning arrays,
where commensurability and enhanced pinning effects
appear \cite{Ray13,Reichhardt07,Reichhardt07a,Reichhardt08}.

\section{Vortex Banding}

\begin{figure}
  \begin{center}
    \includegraphics[width=0.8\columnwidth]{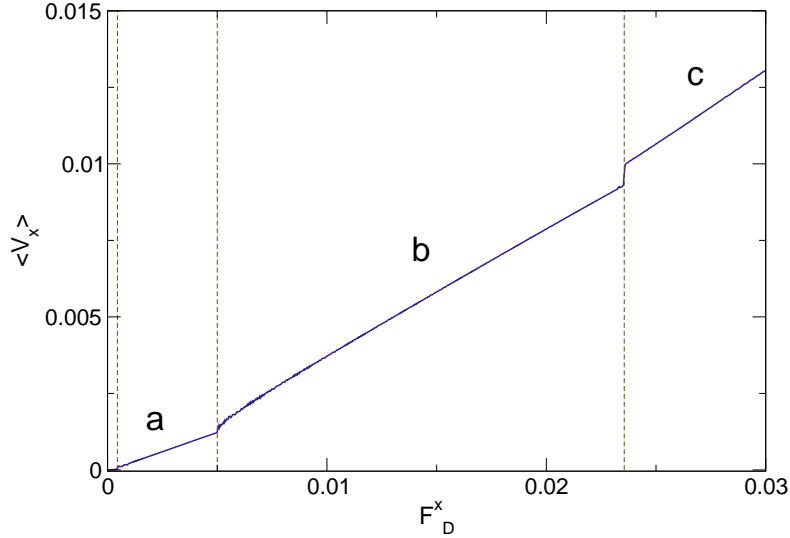}
  \end{center}
  \caption{ The average velocity $\langle V_{x}\rangle$ vs $F^{x}_{D}$ for the
    system in Fig.~\ref{fig:1} at $F^{y}_{D} = 0.1$.
    The leftmost vertical dashed line indicates the transition from a pinned to a
    flowing state and the other two dashed lines mark
    transitions between different shear banded flowing
    states.
    The letters a, b and c correspond to the values of $F^{x}_{D}$
    at which the images in Fig.~\ref{fig:3} were obtained.}
\label{fig:2}
\end{figure}

\begin{figure}
  \begin{center}
    \includegraphics[width=0.8\columnwidth]{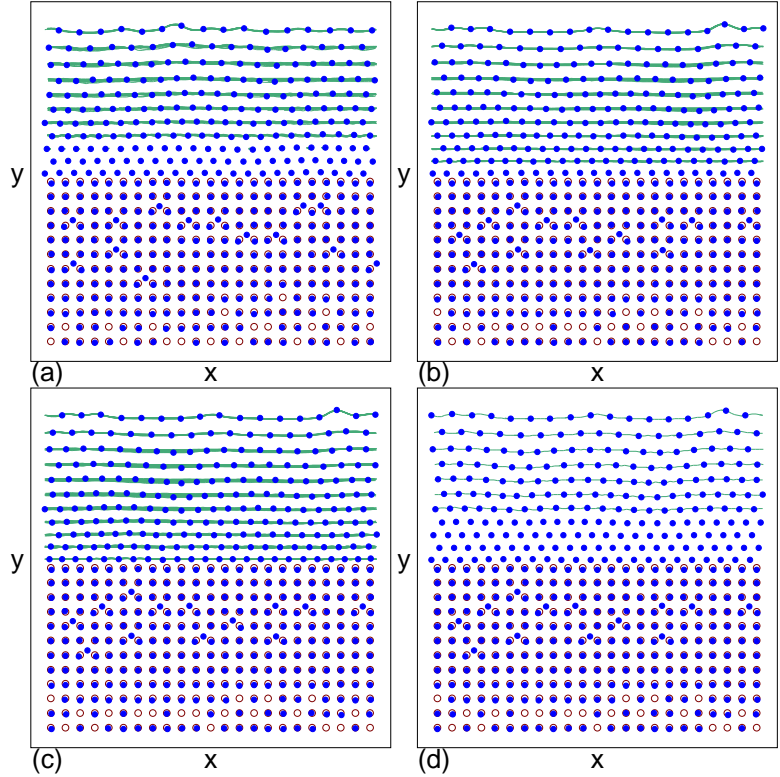}
  \end{center}
  \caption{Pinning site locations (open circles) and vortex positions (dots) and
    trajectories (lines) obtained over a fixed time period for the
    system in Fig.~\ref{fig:2} with $F^y_D=0.1.$
    (a)
    At $F^{x}_{D} = 0.004$,
    there is a band of vortex motion in the upper part of the sample
    and three of the rows in the pin-free channel are immobile.
    (b) At $F^{x}_{D} = 0.0125$,
    there is only one row of immobile vortices in the pin-free channel immediately
    adjacent to the pinning sites.
    (c) At $F^{x}_{D} = 0.025$, all of the vortices in the pin-free channel are flowing.
    (d) In a sample with $F^y_D=0.125$, 
    at $F^{x}_{D} = 0.004$ just above depinning,
    there are four rows of 
    immobile vortices near the edge of the pinned channel.}
\label{fig:3}
\end{figure}

Figure~\ref{fig:1} illustrates
the sample geometry
with $B/B_{\phi} = 2.0$, where
almost all of the pinning sites are occupied.
The arrows indicate the direction of the driving forces
$F^{x}_{D}$ and $F^{y}_{D}$,
which give a net drive magnitude of
$F = \sqrt{(F^{x}_{D})^2 + (F^{y}_{D})^2}$ applied
at an angle $\theta = \tan^{-1}(F^{y}_{D}/F^{x}_{D})$
to the $x$ axis.
In Fig.~\ref{fig:2} we plot
$\langle V_x\rangle$ versus $F_{D}$ for
the system in Fig.~\ref{fig:1} at a fixed $F^{y}_{D} = 0.1$.
The velocity-force curve shows three jumps that
correspond to transitions between different vortex flow patterns.
For $F^{x}_{D} < 0.00085$, the system is in a pinned state and
the vortices in the pin-free channel are trapped by the interactions with the
pinned vortices along the edges of the channel.
For $ 0.00085 \leq F^{x}_{D} < 0.0058$,
shear banded flow occurs as illustrated in Fig.~\ref{fig:3}(a) for $F_{D} = 0.004$,
where there are three rows of immobile vortices just above the
pinned region.
This is followed by an upward jump 
in $\langle V_{x}\rangle$ at $F_{D} = 0.0058$ to
a state in which there is only a single row of immobile vortices in the pin-free
channel,
as shown in Fig.~\ref{fig:3}(b) at $F_{D} = 0.0125$.
This flow state persists over the range
$0.0058 \leq F^{x}_{D} < 0.0208$,
after which
another upward jump in $\langle V_{x}\rangle$ occurs
and all of the vortices in the pin-free channel flow,
as shown in Fig.~\ref{fig:3}(c) at $F^{x}_{D} = 0.025$. 
For increasing perpendicular drive $F^{y}_{D}$,
we find that the number of immobile rows in the pin-free channel increases,
as illustrated in Fig.~\ref{fig:3}(d)
for a sample with
$F_D^y=0.125$ at $F_D^x=0.004$, just above depinning,
where four rows of vortices are motionless.

\begin{figure}
  \begin{center}
    \includegraphics[width=0.8\columnwidth]{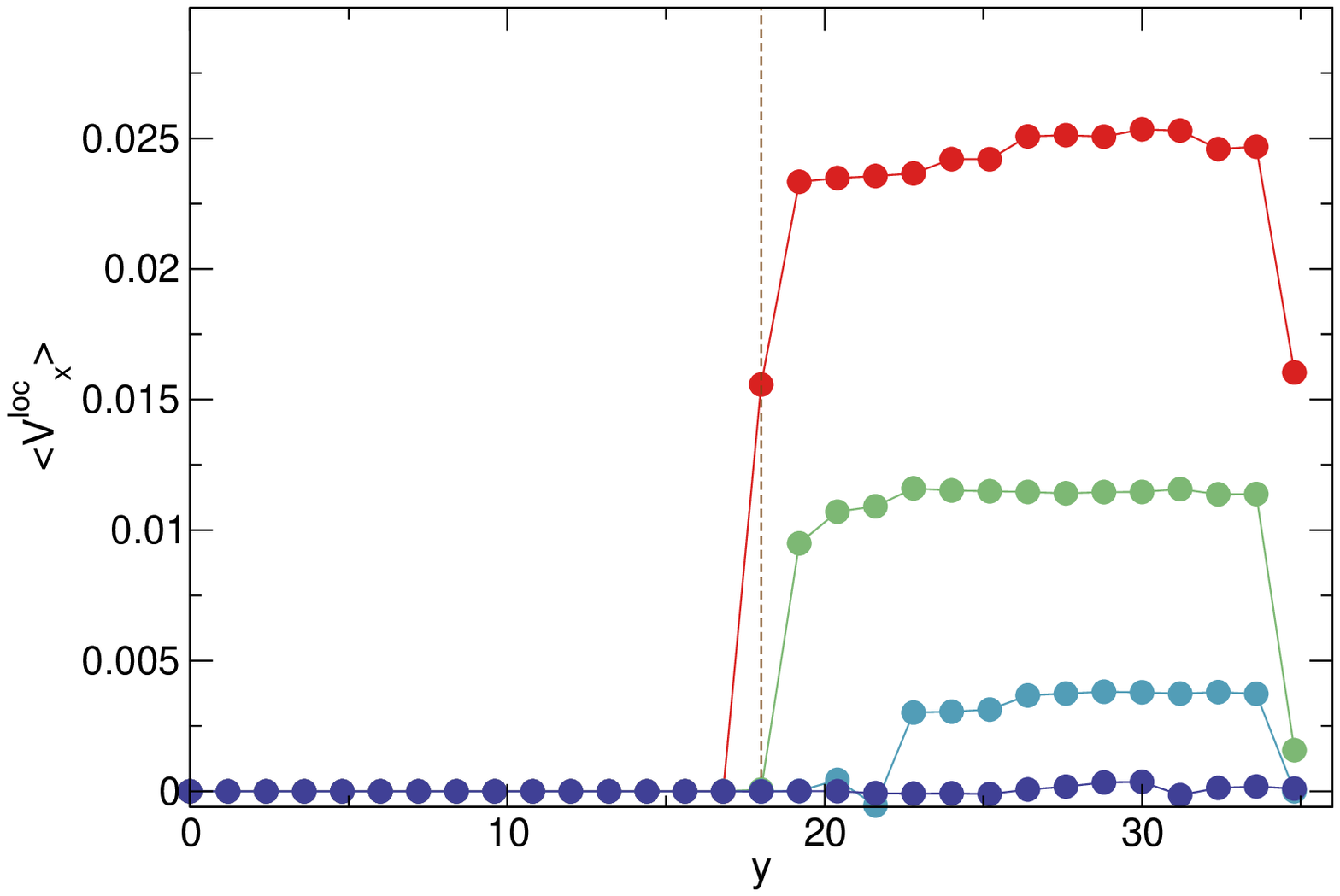}
  \end{center}
  \caption{ The spatial profile of the average local velocity $\langle V_{x}^{\rm loc}\rangle$
    vs $y$ for the system in Fig.~\ref{fig:2} in the four
    different phases.
    The dashed vertical line indicates the edge of the pinned region.
    From bottom to top,
    $F^{x}_{D} = 0.0005$ (dark blue) in the pinned phase, 
    $F^{x}_{D} = 0.004$ (light blue) in a state with three immobile rows in the pin-free
    channel,
    $F^{x}_{D}= 0.0125$ (green) in a state with one immobile row,
    and $F^{x}_{D} = 0.025$ (red) where all the vortices in the pin-free channel are flowing.}
\label{fig:4}
\end{figure}

In Fig.~\ref{fig:4} we plot the spatial average velocity profiles
$\langle V_{x}^{\rm loc}\rangle$ vs $y$ for the system in Fig.~\ref{fig:2}
for the four different phases.
The local average velocity is given by
$\langle V_x^{\rm loc}\rangle(y)=N_{\rm loc}^{-1}\sum_i^{N_v}({\bf v}_i \cdot {\bf \hat x})
\Theta(\epsilon-|r_i^y-y|)$,
where $\epsilon=0.5a$ and $N_{\rm loc}=\sum_i^{N_v}\Theta(\epsilon-|r_i^y-y|)$.
In the pinned state at $F^x_D=0.0005$,
$\langle V_x^{\rm loc}\rangle$ is zero across the entire system,
while at $F^{x}_{D} = 0.004$
where there are three immobile rows in the pin-free channel,
there is a step in the velocity at the edge of the flowing band of vortices.
The velocity is not uniform across the flowing band but is slightly smaller next to
the immobile rows, and we find that the outer moving row undergoes a phase slip
with respect to the remaining moving rows.
At $F_{D} = 0.0125$, where there is one immobile row in the pin-free channel,
the band of finite $\langle V_x^{\rm loc}\rangle$ has extended nearly to the
boundary of the pinning region, and the velocity of the outer moving row is
again lower than that of the rest of the moving band.
At $F^{x}_{D} = 0.025$,
all of the vortices in the pin-free channel are moving
and the band of finite $\langle V_x^{\rm loc}\rangle$ reaches its widest extent,
but the vortex rows closest to the pinning continue to move more slowly than the
other vortices.
The drop in $\langle V_x^{\rm loc}\rangle$ at large $y$ is due to the
presence of the pinned region on the other side of the
periodic boundary.
We note that for all of these values of $F^x_D$,
$\langle V_{y}\rangle$ is zero.

\begin{figure}
  \begin{center}
    \includegraphics[width=0.8\columnwidth]{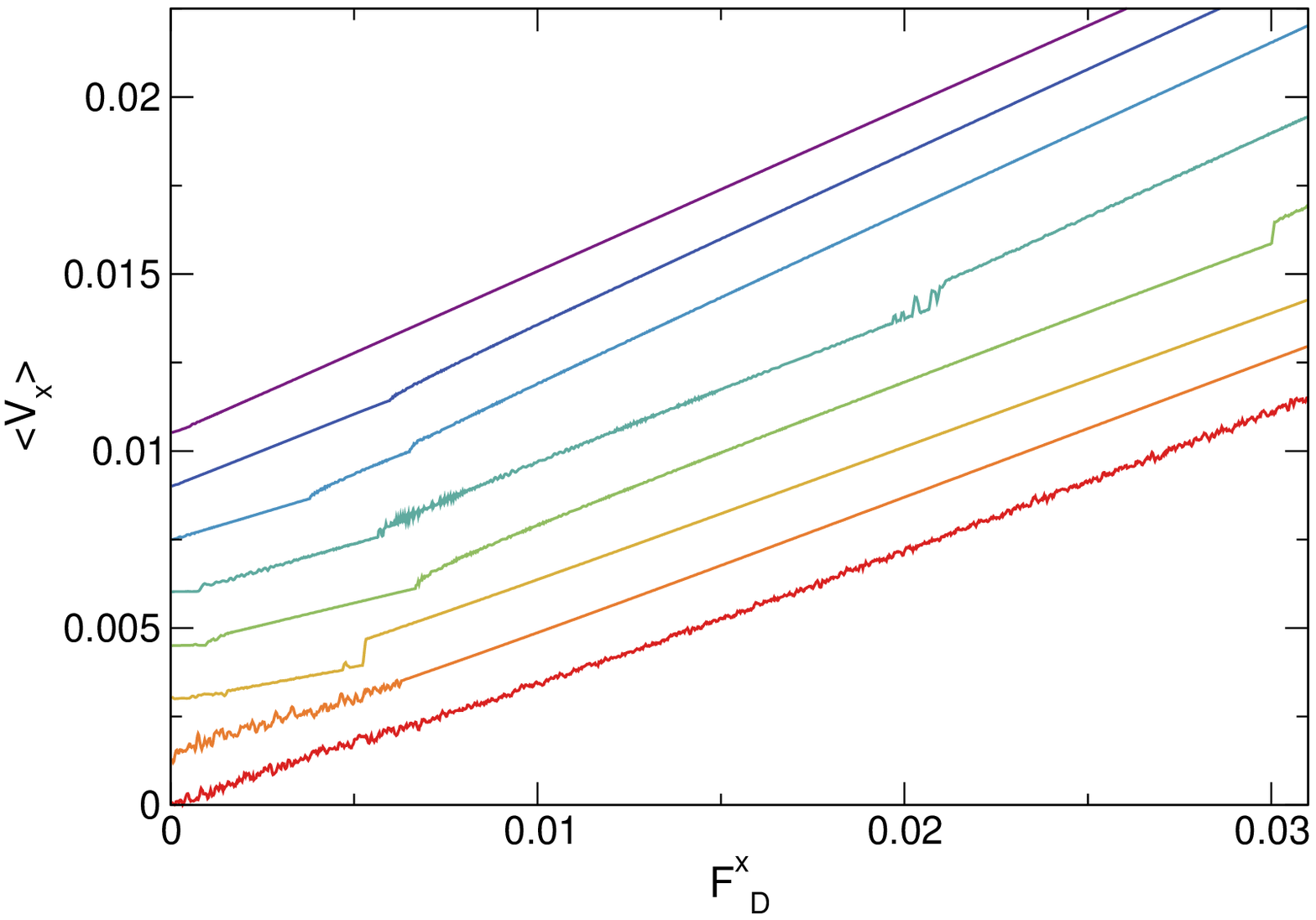}
  \end{center}
\caption{ 
  $\langle V_{x}\rangle$ vs $F^{x}_{D}$ for
  samples with, from top to bottom, $F^{D}_{y} = 0.025$ (purple),
  $0.0375$ (dark blue),
  $0.05$ (light blue),
  $0.1$ (dark green),
  $0.125$ (light green),
  $0.15$ (yellow), $0.1625$ (orange),
  and $0.175$ (red).
  For $F^{x}_{D} = 0.175$, we find
  a disordered flow phase R
  where there is motion in both the $x$ and $y$ directions,
  as illustrated in Fig.~\ref{fig:6}(b).
  The curves have been shifted vertically for clarity.
}
\label{fig:5}
\end{figure}

\begin{figure}
  \begin{center}
    \includegraphics[width=0.8\columnwidth]{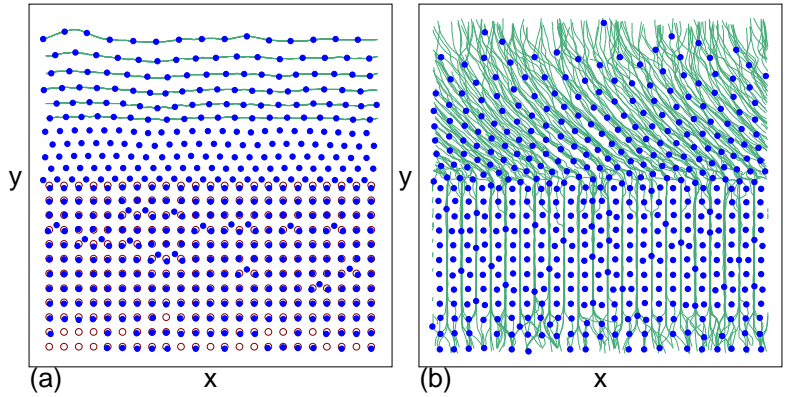}
  \end{center}
  \caption{ Pinning site locations (open circles) and vortex positions (dots) and
    trajectories (lines) obtained over a fixed time period.
    (a) Phase V at $F^{x}_{D} = 0.004$ and $F^{y}_{D} = 0.15$,
    where there are five immobile vortex rows in the pin-free channel.
    (b) The random guided phase R at $F^{y}_{D} = 0.175$ and $F^{x}_{D} = 0.015$,
    where motion in the pin-free channel is in both the $x$ and $y$ directions,
    while the motion in the pinned region is strictly in the $y$ direction.
    }
\label{fig:6}
\end{figure}

We now introduce some terminology for the different states.
We denote the pinned state as phase P, while
in phase M, all of the rows in the pin-free channel are flowing.
In phase I, there is one immobile row of  vortices in the pin-free channel,
in phase III, there are three immobile rows, and so forth.
For the parameters considered here, we have not observed a situation in which
there are two immobile rows in the pin-free channel.
In Fig.~\ref{fig:5} we plot a series of $\langle V_{x}\rangle$
versus $F^{x}_{D}$ curves for $F^{D}_{y} = 0.025$, 
$0.0375$, $0.05$, $0.1$, $0.125$, 
$0.15$, $0.1625$, and $0.175$.
At $F^{y}_{D} = 0.025$,
there is only a single transition from the pinned
phase P to phase M in which all the vortices in the pin-free channel are moving.
For $F^{y}_{D} = 0.0375$, the pinned phase P is followed first by
phase III and then by phase M,
while for $F^{y}_{D} = 0.1$,
the sequence of states is phase P to phase III to phase I to phase M.
When $F^{y}_{D} = 0.125$, the system transitions from
the pinned phase P to phase IV in which there are four immobile vortex rows,
followed by phase I and then by phase M, which appears
near $F^{x}_{D} = 0.3$.
At $F^{y}_{D} = 0.15$, we find a transition from phase P to
phase V,
where there are five pinned rows of interstitial vortices as illustrated
in Fig.~\ref{fig:6}(a).
A transition to phase I occurs
at a much higher drive of $F^{x}_{D} = 0.055$.
For $F^{y}_{D} = 0.1625$, the pinned phase P is lost
and the system
is initially
in a random guided flow, where the particles
flow in both the $x$ and $y$ directions in the pin-free channel but in only the
$y$ direction in the pinned region.
This state is illustrated in Fig.~\ref{fig:6}(b)
for a system with $F^y_D=0.175$ and $F^x_D=0.015$.
For sufficiently large $F_D^x$, the $F^{y}_{D} = 0.1625$ system
can transition from phase R to phase I,
but when $F^{y}_{D} > 0.1625$,
the system remains in phase R and there are never immobile vortex
rows in the pin-free channel.

\begin{figure}
  \begin{center}
    \includegraphics[width=0.8\columnwidth]{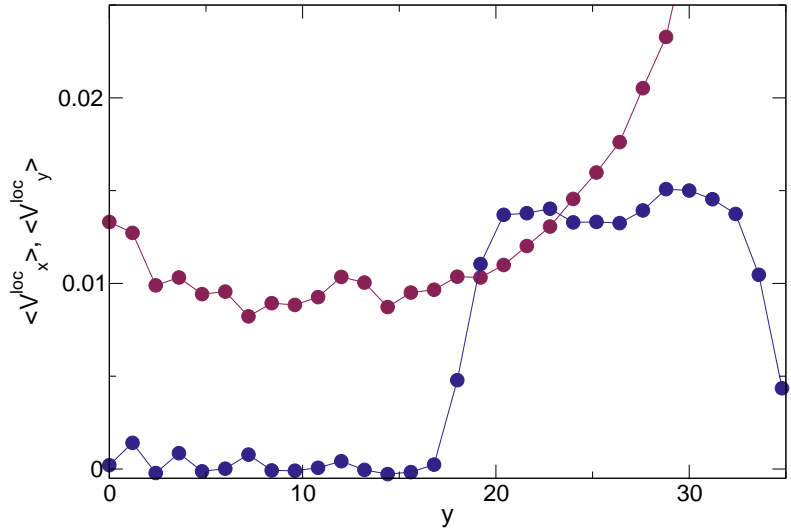}
  \end{center}
\caption{ 
  The spatial profile of the local velocities $\langle V_x^{\rm loc}\rangle$ (blue) and
  $\langle V_y^{\rm loc}\rangle$ (red)
  vs $y$ for the phase R flow in the system
  from Fig.~\ref{fig:6}(b). Within the pinned region,
  the flow is strictly in the $y$ direction,
  while in the pin-free channel,
  the flow is in both the $x$ and $y$ directions.
  For clarity, the $\langle V^{\rm loc}_y\rangle$ curve has been scaled down by
  a factor of two.
}
\label{fig:7}
\end{figure}

To further characterize the phase R flow,
in Fig.~\ref{fig:7} we plot the spatial velocity profiles
$\langle V^{\rm loc}_x\rangle$ and $\langle V^{\rm loc}_y\rangle$ versus $y$
for the flow pattern illustrated in Fig.~\ref{fig:6}(b).
Within the pinned region, the vortices are moving strictly in the $y$-direction,
while in the pin-free channel,
the vortices are moving in both the $x$ and $y$ directions.
We note that the $\langle V^{\rm loc}_{y}\rangle$ curve has been reduced by 
a factor of two for clarity.

\begin{figure}
  \begin{center}
    \includegraphics[width=0.8\columnwidth]{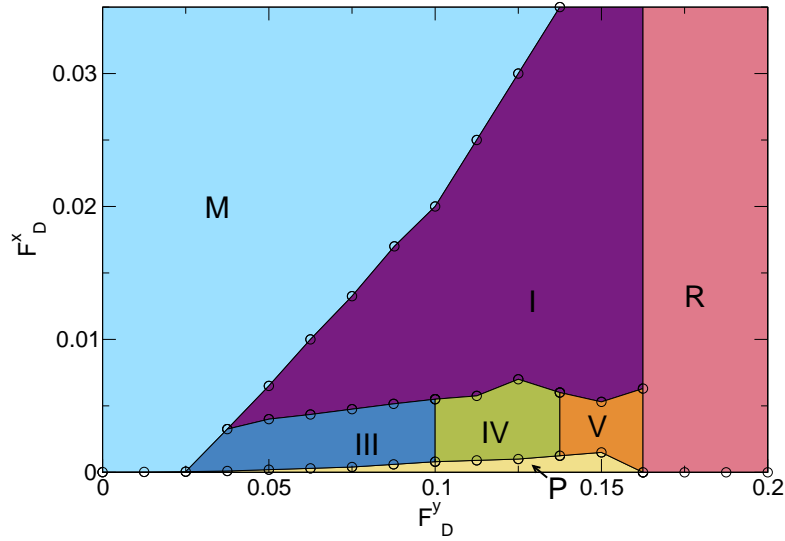}
  \end{center}
  \caption{ The dynamic phase diagram as a function of $F^{x}_{D}$
    vs $F^{y}_{D}$ for the system in Fig.~\ref{fig:6}.
    P: pinned phase (yellow);
    V: five immobile vortex rows in the pin-free channel (orange);
    IV: four immobile vortex rows (green);
    III: three immobile vortex rows (dark blue);
    I: one immobile vortex row (purple);
    M: motion of all vortices in the pin-free channel (light blue);
    R: random guided flow in the entire sample (pink).
  }
\label{fig:8}
\end{figure}

In Fig.~\ref{fig:8} we plot the dynamic phase diagram as a function of
$F^{x}_{D}$ versus $F^{y}_{D}$ for the
system from Fig.~\ref{fig:6},
highlighting the pinned phase P,
phases I, III, IV, and V,
phase M in which all vortices in the pin-free channel are moving,
and the random guided flow phase R.
We note that for finer intervals of $F^{y}_{D}$ or
a larger system,
there will likely be
additional phases corresponding to six, seven,
and higher numbers of immobile vortex rows
in the pin-free channel.
For higher values of $F^{x}_{D}$
than what is considered here, there
are additional phases that occur
when the vortices at the pinning sites begin to depin,
which 
is beyond the scope of this work.
We also note that in general we do not observe a continuous velocity gradient
in our system.
This is probably a result of
the triangular ordering of the vortices within the pin-free channel, which
causes the
system to
behave like a sheared triangular solid breaking along its easy shear direction.
If the system were in a liquid or amorphous state,
it could be possible to obtain
a continuous velocity gradient with no well-defined shear bands.

\section{Shearing Dynamics for Perpendicular Driving and Field Effect Transistor}

\begin{figure}
  \begin{center}
    \includegraphics[width=0.8\columnwidth]{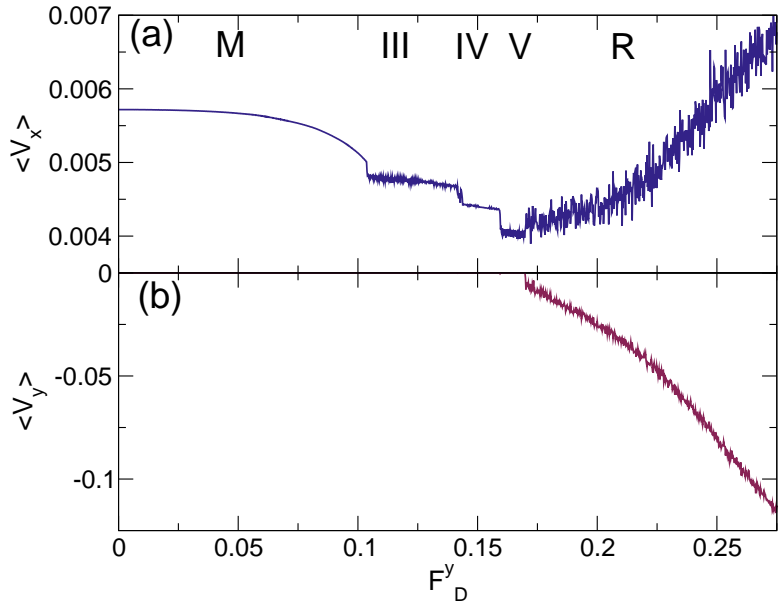}
  \end{center}
  \caption{(a) $\langle V_{x}\rangle$ vs $F^{y}_{D}$ for the system in
    Fig.~\ref{fig:2} at a constant $F^{x}_{D} = 0.0125$, 
    showing a series of drops corresponding to transitions
    among phases M, III, IV, V, and R.
    (b) The corresponding $\langle V_{y}\rangle$ vs $F^{y}_{D}$,
    showing that $\langle V_y\rangle$ only becomes finite in phase R. }
\label{fig:9}
\end{figure}

We next consider the case where the driving
force along the $x$ or parallel direction is fixed
while the drive along the $y$ or perpendicular direction increases.
Although the net force
$F=\sqrt{{F^x_D}^2+{F^y_D}^2}$ is increasing,
the net velocity 
$\langle V\rangle = \sqrt{\langle V_{x}\rangle^2 + \langle V_{y}\rangle^2}$
generally decreases in a series of steps as more rows of vortices become
immobile.
In Fig.~\ref{fig:9}
we plot $\langle V_{x}\rangle$ and $\langle V_{y}\rangle$
versus $F^{y}_{D}$ for the same system in Fig.~\ref{fig:2} 
at fixed $F^{x}_{D} = 0.0125$.
When $F^{y}_{D} = 0$, the system is in phase M, with all the vortices
in the pin-free channel flowing.
As $F^y_D$ increases,
$\langle V_{x}\rangle$ gradually decreases,
with a downward jump near $F^{y}_{D} = 0.105$
marking the transition into phase III.
This is followed by 
two more drops in $\langle V_x\rangle$ at the
transitions into phases IV and V,
and at $F^y_D=0.17$,
the system enters phase R,
where $\langle V_{y}\rangle$ becomes finite
and the velocities
in both directions show large fluctuations.
In phase R,
$\langle V_x\rangle$ begins to increase with
increasing $F^{y}_{D}$.
For other values of $F^x_D$, we observe a similar
trend with drops
in $\langle V_{x}\rangle$ marking the transitions among different phases;
however, for this pinning density,
we do not find a phase in which
$\langle V_{x}\rangle$ goes to zero.
For denser pinning, it is more difficult for the vortices to enter the pinned
region, and it is possible to observe a reentrant pinning effect.

\begin{figure}
  \begin{center}
    \includegraphics[width=0.8\columnwidth]{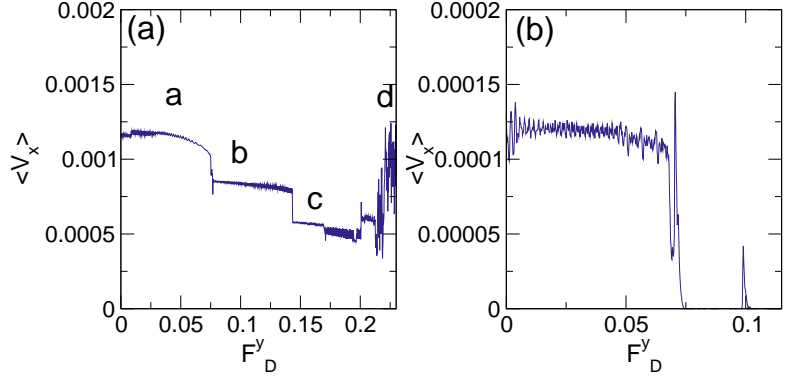}
  \end{center}
  \caption{ (a) $\langle V_{x}\rangle$ vs $F^{y}_{D}$
    at fixed $F^{x}_{D} = 0.0025$ for a system with a
    higher pin density of $B_\phi^{\rm complete}=0.6/\lambda^2$.
    The letters a to d indicate the drives at which the images
    in Fig.~\ref{fig:11} were obtained.
    (b) $\langle V_x\rangle$ vs $F^y_D$ for the same system
    at fixed $F^{x}_{D} = 0.00025$, 
    where there is a transition into a pinned
    state with $\langle V_{x}\rangle = 0$.  }
\label{fig:10}
\end{figure}

\begin{figure}
  \begin{center}
    \includegraphics[width=0.8\columnwidth]{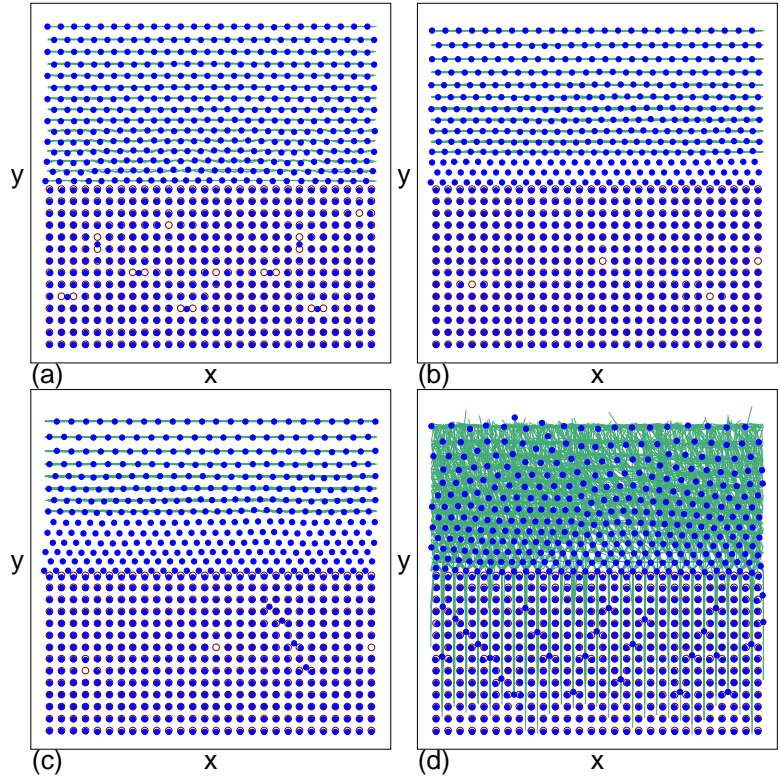}
  \end{center}
  \caption{Pinning site locations (open circles) and vortex positions (dots) and
    trajectories (lines) obtained over a fixed time period
    for the system in Fig.~\ref{fig:9}.
    (a) At $F^{y}_{D}= 0.05$, all the vortices in the pin-free channel are flowing.
    (b) At $F^{D}_{y} = 0.1$,
    there are three immobile vortex rows in the pin-free channel.
    (c) At $F^{y}_{D} = 0.15$,
    there are six immobile vortex rows in the pin-free channel.
    (d) Phase R at $F^{y}_{D} = 0.225$. }
\label{fig:11}
\end{figure}

In Fig.~\ref{fig:10}(a) we plot
$\langle V_{x}\rangle$ versus $F^{y}_{D}$ for a system with
fixed $F^{x}_{D} = 0.0025$ but denser
pinning of $B_\phi^{\rm complete}=0.6/\lambda^2$ at the same filling of
$B/B_\phi=2.0$.
For $0 < F^{y}_{D} < 0.075$, the system is in phase M
with all the vortices in the pin-free channel flowing,
as illustrated
in Fig.~\ref{fig:11}(a) at $F^{y}_{D} = 0.05$.
There is a jump down in $\langle V_x\rangle$ at the transition to
a phase in which there are three immobile rows of vortices in the
pin-free channel, which appears over the range
$0.075 < F^{y}_{D} < 0.1435$
and is illustrated in Fig.~\ref{fig:11}(b) for $F^{y}_{D} = 0.1$.
When $0.1435 < F^{y}_{D} < 0.2$, there
are
six immobile vortex rows
in the pin-free channel, as shown in Fig.~\ref{fig:11}(c) at $F^{y}_{D} = 0.15$.
For $0.2 < F^{y}_{D} < 0.22$, 
an upward jump in
$\langle V_{x}\rangle$ is accompanied by a transition to a state
with  five immobile vortex rows in the pin-free channel.
At higher $F^y_D$, the system enters region R,
as illustrated in Fig.~\ref{fig:11}(d) at $F^{y}_{D} = 0.225$, where
the vortices only flow along the $y$ direction within
the pinning region.

In Fig.~\ref{fig:10}(b) we plot
$\langle V_{x}\rangle$ versus $F^{y}_{D}$
for the same system as in Fig.~\ref{fig:10}(a) but at a smaller value of 
$F^{D}_{x} = 0.0025$.
Here the system is initially
in phase M before transitioning to a pinned state
with $\langle V_{x}\rangle = 0$.
Eventually, when $F^{y}_{D}$ is large enough (not shown),
the system enters phase R flow.
The
sharp drop down in the velocity at the onset of phase P
is similar to what is observed in a field effect transition,
where an increase in driving 
can cause a sharp transition in the current flow,
or the vortex flow in this case.
In general, as the density of pinning sites increases,
the range of driving over which
this field effect transistor
phenomenon occurs becomes wider.

\section{Summary}
We numerally examine vortices in a system with a square pinning array
where half of the pinning sites have been removed to form a pin-free channel.
When a drive is applied at an angle with respect to the channel,
we find 
that
the vortices in the
channel depin in a series of transitions
among different shear banded flow
phases.
The vortex velocity
exhibits a sequence of steps corresponding to
the quantization of the number of rows of moving vortices. 
As the component of the driving force parallel to the channel increases,
the depinning of individual rows of vortices produces
steps in the velocity-force curve.
When the component of the driving force perpendicular to the channel increases,
the number of immobile vortex rows in the pin-free channel also increases.
A gradient in the velocity of the moving vortices appears in the pin-free channel,
with vortices close to the edge of the channel moving more slowly than those in the
center of the channel.
When the perpendicular drive is large enough, 
there is
a transition to a state in which the vortices move
both
parallel and perpendicular to the pin-free channel;
however, the vortices
within the pinned part of the sample move only
perpendicular to the orientation of the pin-free channel.
If the parallel driving force is fixed and the perpendicular driving force is
increased,
we find
a series
of downward steps in the longitudinal velocity
as the number of immobile vortex rows increases,
and in some cases it is even possible
for the increasing
perpendicular drive
to produce a complete suppression of vortex motion
in a field effect transistor phenomenon.
In addition to
vortices, our results should be general to
other particle-based systems with inhomogeneous pinning,
such as colloids and skyrmions.  

\ack
This work was supported by the US Department of Energy through
the Los Alamos National Laboratory.  Los Alamos National Laboratory is
operated by Triad National Security, LLC, for the National Nuclear Security
Administration of the U. S. Department of Energy (Contract No. 892333218NCA000001).

\section*{References}
\bibliographystyle{iopart-num}
\bibliography{mybib}

\providecommand{\newblock}{}
\begin{thebibliography}{10}
\expandafter\ifx\csname url\endcsname\relax
  \def\url#1{{\tt #1}}\fi
\expandafter\ifx\csname urlprefix\endcsname\relax\def\urlprefix{URL }\fi
\providecommand{\eprint}[2][]{\url{#2}}

\bibitem{Beer09}
Beer F, Johnston R, Dewolf J and Mazurek D 2009 {\em Mechanics of Materials\/}
  (McGraw-Hill, New York)

\bibitem{Argon79}
Argon A~S 1979 {\em Acta Metallurg.\/} {\bf 27} 47--58

\bibitem{Regev13}
Regev I, Lookman T and Reichhardt C 2013 {\em Phys. Rev. E\/} {\bf 88}(6)
  062401

\bibitem{Cohen06}
Cohen I, Davidovitch B, Schofield A~B, Brenner M~P and Weitz D~A 2006 {\em
  Phys. Rev. Lett.\/} {\bf 97}(21) 215502

\bibitem{AlixWilliams18}
Alix-Williams D~D and Falk M~L 2018 {\em Phys. Rev. E\/} {\bf 98}(5) 053002

\bibitem{Fisher98}
Fisher D~S 1998 {\em Phys. Rep.\/} {\bf 301}(1-3) 113--150

\bibitem{Reichhardt17}
Reichhardt C and Reichhardt C~J~O 2017 {\em Rep. Prog. Phys.\/} {\bf 80}(2)
  026501

\bibitem{Lopez99}
L\'opez D, Kwok W~K, Safar H, Olsson R~J, Petrean A~M, Paulius L and Crabtree
  G~W 1999 {\em Phys. Rev. Lett.\/} {\bf 82}(6) 1277--1280

\bibitem{Benetatos02}
Benetatos P and Marchetti M~C 2002 {\em Phys. Rev. B\/} {\bf 65}(13) 134517

\bibitem{Miguel03}
Miguel M~C and Zapperi S 2003 {\em Nature Mater.\/} {\bf 2}(7) 477--481

\bibitem{Furukawa06}
Furukawa A and Nisikawa Y 2006 {\em Phys. Rev. B\/} {\bf 73}(6) 064511

\bibitem{Okuma09}
Okuma S, Yamazaki Y and Kokubo N 2009 {\em Phys. Rev. B\/} {\bf 80}(22) 220501

\bibitem{Cabral16}
Cabral L~R~E, de~Aquino B~R~C~H~T, Silva C~C~d~S, Milo\ifmmode \check{s}\else
  \v{s}\fi{}evi\ifmmode~\acute{c}\else \'{c}\fi{} M~V and Peeters F~M 2016 {\em
  Phys. Rev. B\/} {\bf 93}(1) 014515

\bibitem{Miguel11}
Miguel M~C, Mughal A and Zapperi S 2011 {\em Phys. Rev. Lett.\/} {\bf 106}(24)
  245501

\bibitem{Lin09}
Lin N~S, Misko V~R and Peeters F~M 2009 {\em Phys. Rev. Lett.\/} {\bf 102}(19)
  197003

\bibitem{Kawamura15}
Kawamura Y, Matsumura Y, Yamazaki Y, Kaneko S, Kokubo N and Okuma S 2015 {\em
  Supercond. Sci. Technol.\/} {\bf 28}(4) 045002

\bibitem{Rosen13}
Rosen Y~J, Gu\'enon S and Schuller I~K 2013 {\em Phys. Rev. B\/} {\bf 88}(17)
  174511

\bibitem{Reichhardt10}
Reichhardt C and Reichhardt C~J~O 2010 {\em Phys. Rev. B\/} {\bf 81}(10) 100506

\bibitem{Marchetti99}
Marchetti M~C and Nelson D~R 1999 {\em Phys. Rev. B\/} {\bf 59}(21)
  13624--13627

\bibitem{Marchetti00}
Marchetti M~C and Nelson D~R 2000 {\em Physica C\/} {\bf 330}(3-4) 105--129

\bibitem{Buttinoni15}
Buttinoni I, Zell Z~A, Squires T~M and Isa L 2015 {\em Soft Matter\/} {\bf
  11}(42) 8313--8321

\bibitem{Williams16}
Williams I, Oguz E~C, Speck T, Bartlett P, L{\" o}wen H and Royall C~P 2016
  {\em Nature Phys.\/} {\bf 12}(1) 98--U134

\bibitem{OrtizAmbriz18}
Ortiz-Ambriz A, Gerloff S, Klapp S~H~L, Ortin J and Tierno P 2018 {\em Soft
  Matter\/} {\bf 14}(24) 5121--5129

\bibitem{Baert95}
Baert M, Metlushko V~V, Jonckheere R, Moshchalkov V~V and Bruynseraede Y 1995
  {\em Phys. Rev. Lett.\/} {\bf 74} 3269--3272

\bibitem{Martin97}
Mart\'{\i}n J~I, V\'elez M, Nogu\'es J and Schuller I~K 1997 {\em Phys. Rev.
  Lett.\/} {\bf 79}(10) 1929--1932

\bibitem{Trastoy14}
Trastoy J, Malnou M, Ulysse C, Bernard R, Bergeal N, Faini G, Lesueur J,
  Briatico J and Villegas J~E 2014 {\em Nature Nanotechnol.\/} {\bf 9}(9)
  710--715

\bibitem{Wang13}
Wang Y~L, Latimer M~L, Xiao Z~L, Divan R, Ocola L~E, Crabtree G~W and Kwok W~K
  2013 {\em Phys. Rev. B\/} {\bf 87}(22) 220501

\bibitem{Banerjee03}
Banerjee S~S, Soibel A, Myasoedov Y, Rappaport M, Zeldov E, Menghini M, Fasano
  Y, de~la Cruz F, van~der Beek C~J, Konczykowski M and Tamegai T 2003 {\em
  Phys. Rev. Lett.\/} {\bf 90}(8) 087004

\bibitem{Besseling99}
Besseling R, Niggebrugge R and Kes P~H 1999 {\em Phys. Rev. Lett.\/} {\bf
  82}(15) 3144--3147

\bibitem{Guillamon14}
Guillam{\' o}n I, C{\' o}rdoba R, Ses{\' e} J, De~Teresa J~M, Ibarra M~R, Viera
  S and Suderow H 2014 {\em Nature Phys.\/} {\bf 10}(11) 851--856

\bibitem{Dobrovolskiy19}
Dobrovolskiy O, Bevz V, Begun E, Sachser R, Vovk R and Huth M 2019 {\em Phys.
  Rev. Applied\/} {\bf 11}(5) 054064

\bibitem{Dobrovolskiy17}
Dobrovolskiy O~V, Shklovskij V~A, Hanefeld M, Z{\" o}rb M, K{\" o}hs L and Huth
  M 2017 {\em Supercond. Sci. Technol.\/} {\bf 30}(8) 085002

\bibitem{Dobrovolskiy16}
Dobrovolskiy O~V, Hanefeld M, Z{\" o}rb M, Huth M and Shklovskij V~A 2016 {\em
  Supercond. Sci. Technol.\/} {\bf 29}(6) 065009

\bibitem{Ray13}
Ray D, Olson~Reichhardt C~J, Jank\'o B and Reichhardt C 2013 {\em Phys. Rev.
  Lett.\/} {\bf 110}(26) 267001

\bibitem{Guenon13}
Guenon S, Rosen Y~J, Basaran A~C and Schuller I~K 2013 {\em Appl. Phys.
  Lett.\/} {\bf 102}(25) 252602

\bibitem{Reichhardt07}
Reichhardt C and Reichhardt C~J~O 2007 {\em Phys. Rev. B\/} {\bf 76}(6) 064523

\bibitem{Reichhardt07a}
Reichhardt C and Reichhardt C~J~O 2007 {\em Phys. Rev. B\/} {\bf 76}(9) 094512

\bibitem{Reichhardt08}
Reichhardt C and Olson~Reichhardt C~J 2008 {\em Phys. Rev. B\/} {\bf 78}(18)
  180507

\end{thebibliography}
\end{document}